\pdfoutput=1

\documentclass[reprint,amsmath,amssymb,floatfix,aps,longbibliography]{revtex4-1}
\usepackage{graphics,graphicx,float}

\usepackage[pdftex,
            pdfauthor={Steven A. Frank},
            pdftitle={},
            pdfsubject={Natural selection. I. Variable environments and uncertain returns on investment}]{hyperref} 

\def\citeyear{\citep}
\def\autocite{\citep}

\newcommand{\Rbar}{\overline{R}}
\DeclareMathOperator{\cov}{cov}
\DeclareMathOperator{\var}{var}
\DeclareMathOperator{\E}{E}

\newcommand{\Gm}{\mu}
\newcommand{\Gd}{\delta}
\newcommand{\Gs}{\sigma}
\newcommand{\GD}{\Delta}
\newcommand{\Gr}{\rho}
\newcommand{\Ge}{\epsilon}
\newcommand{\Gg}{\gamma}

\newcommand{\Eq}[1]{Eq.~(\ref{eq:#1})}

\newcommand{\Fig}[1]{Fig.~\ref{fig:#1}}

\newcommand{\boldrule}{\hrule height 1.2pt}
\newcommand{\noterule}{\medskip\boldrule\medskip}	

\newcount\BoxNum \BoxNum 1\relax
\makeatletter
\newcommand{\boxlabel}[1]{%
  \protected@write \@auxout {}{\string \newlabel {box:#1}{{\the\BoxNum}{\thepage}{\noexpand\relax}%
  	{\@ifundefined{hyper@@anchor}{\relax}{box.\the\BoxNum}}%
  	{}}}%
  \@ifundefined{hyper@@anchor}{}{\hypertarget{box.\the\BoxNum}{}}%
  \advance\BoxNum 1\relax}
\makeatother
\newcommand{\Boxx}[1]{Box~\ref{box:#1}}
\newcommand{\BoxLabel}{Box~\the\BoxNum}

\begin{document}

\title{Natural selection. I. Variable environments and uncertain returns on investment}

\author{Steven A.\ Frank}
\email[email: ]{safrank@uci.edu}
\homepage[homepage: ]{http://stevefrank.org}
\affiliation{Department of Ecology and Evolutionary Biology, University of California, Irvine, CA 92697--2525  USA}

\begin{abstract}

Many studies have analyzed how variability in reproductive success affects fitness.  However, each study tends to focus on a particular problem, leaving unclear the overall structure of variability in populations.  This fractured conceptual framework often causes particular applications to be incomplete or improperly analyzed.  In this paper, I present a concise introduction to the two key aspects of the theory.  First, all measures of fitness ultimately arise from the relative comparison of the reproductive success of individuals or genotypes with the average reproductive success in the population.  That relative measure creates a diminishing relation between reproductive success and fitness.  Diminishing returns reduce fitness in proportion to variability in reproductive success.  The relative measurement of success also induces a frequency dependence that favors rare types.  Second, variability in populations has a hierarchical structure.  Variable success in different traits of an individual affects that individual's variation in reproduction.  Correlation between different individuals' reproduction affects variation in the aggregate success of particular alleles across the population.  One must consider the hierarchical structure of variability in relation to different consequences of temporal, spatial, and developmental variability.  Although a complete analysis of variability has many separate parts, this simple framework allows one to see the structure of the whole and to place particular problems in their proper relation to the general theory.  The biological understanding of relative success and the hierarchical structure of variability in populations may also contribute to a deeper economic theory of returns under uncertainty\footnote{\href{http://dx.doi.org/10.1111/j.1420-9101.2011.02378.x}{doi:\ 10.1111/j.1420-9101.2011.02378.x} in \textit{J. Evol. Biol.}}\footnote{Part of the Topics in Natural Selection series. See \Boxx{preface}.}.
\end{abstract}

\maketitle

\section{Introduction}

Natural selection favors traits that enhance fitness.  But how does one measure fitness?  Several studies have shown that it is not just average reproductive success that matters. Variation in reproductive success also plays an important role in determining long-term evolutionary trends.  To understand the basic notions of fitness and evolutionary change by natural selection, one must understand the particular consequences of different kinds of variation.  

The literature on variation splits into two groups.  On one side, bits of folk wisdom dominate thinking. The slogan that natural selection maximizes geometric mean fitness is one example.  Such folk wisdom is true in special cases.  But as a guiding principle, the simple geometric mean slogan misleads as often as it helps.  

On the other side, a technically demanding specialist literature divides into numerous distinct ways of framing the problem.  Each technical expression emphasizes a particular aspect of variation, refining unique examples at the expense of providing a coherent view of the whole.  

This paper provides a tutorial on the different kinds of variation and their evolutionary consequences.  I emphasize simple examples to develop understanding of temporal, spatial, developmental, and trait variation.  Each type of variation was originally studied as a separate problem.  In this tutorial, I follow \textcite{frank90evolution}, who showed that these seemingly different types of variation can be understood in a unified way.  The unified framework arises from two steps.  

First, it is relative reproduction that matters.  Only those traits associated with relatively greater success than average increase over time.  Relative measures of success induce diminishing returns: a doubling of reproduction provides less than a doubling of relative success \autocite{gillespie77natural,frank90evolution}.  With diminishing returns, increasing variation in reproductive success reduces fitness.

Second, the different types of variation can be expressed as different levels within a unified hierarchy \autocite{frank90evolution}.  Variable success in different traits of an individual affects that individual's variation in reproduction.  Correlation between different individuals' reproduction affects variation in the aggregate success of particular alleles across the population.  Temporal, spatial, and developmental variation affect the way in which individual variations combine to determine the overall variability in the number of copies produced by a particular allele.  

I also discuss the relation of economic theories of risk and uncertainty to evolutionary theories of variability.

\section{Relative success induces diminishing returns}

The success of genes and of traits must ultimately be measured by their relative frequency in a population.  The calculation of relative frequency leads to surprising consequences when there is variability \autocite{gillespie77natural,frank90evolution,orr07absolute}.  

To illustrate the problem, consider two alternative types in a population, $A_1$ and $A_2$.  I will often refer to the alternative types as alleles at a genetic locus.  However, 

\begin{figure}[H]
\begin{minipage}{\hsize}
\parindent=15pt
\noterule
{\bf \noindent\BoxLabel. Topics in the theory of natural selection}
\noterule
This article is the first in a series on natural selection.  Although the theory of natural selection is simple, it remains endlessly contentious and difficult to apply.  My goal is to make more accessible the concepts that are so important, yet either mostly unknown or widely misunderstood.  I write in a nontechnical style, showing the key equations and results rather than providing full derivations or discussions of mathematical problems.  Boxes list technical issues and brief summaries of the literature.
\noterule
\end{minipage}
\end{figure}
\boxlabel{preface}

\noindent the same analysis would apply to any heritable alternative types in a population that have the same essential properties as alleles.

Some simple notation helps to express the argument.  Each definition uses subscripts to associate with the alternative alleles, $A_1$ and $A_2$, respectively. Define $q_1$ and $q_2$ as the allele frequencies, such that $q_1+q_2=1$.  Let $R_1$ and $R_2$ measure reproductive success, the average number of descendant copies produced by each parental allele. The average reproductive success in the population is $\Rbar=q_1R_1+q_2R_2$. The success of individual parental copies has a random component.  Thus, all of the measures of reproductive success fluctuate randomly.  Throughout, the unqualified words \textit{average} and \textit{mean} refer to the arithmetic average.

The frequency of $A_1$ after one round of reproduction is 
\begin{equation}\label{eq:freq}
  q_1'=q_1(R_1/\Rbar)= q_1F_1,
\end{equation}
where $F$ measures relative success.  I use the word \textit{fitness} for relative success. This equation shows that fitness ultimately determines the success of an allele.  Reproductive success, $R$, influences fitness.  But the key relationship between reproductive success and fitness is mediated through the definition for fitness
\begin{equation}\label{eq:fitness}
  F_1=R_1/\Rbar=\frac{R_1}{q_1R_1+q_2R_2}.
\end{equation}

\begin{figure}[t]
\includegraphics[width=0.85\hsize]{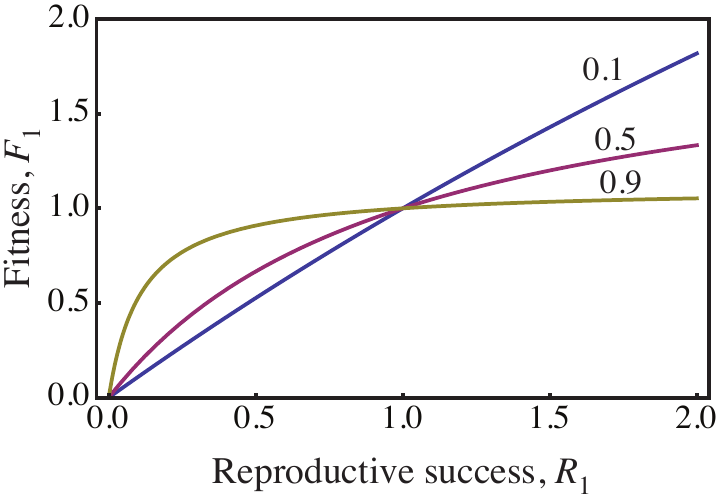}
\caption{The curvature of fitness versus reproductive success depends on allele frequency. The plots here illustrate \Eq{fitness}.  The numbers above each curve show $q_1$, the frequency of the allele for which the relationship is plotted. If $q_1=0.1$, the relationship follows the upper curve for allele $A_1$; the lower curve can then be interpreted as the relationship for allele $A_2$ with frequency $q_2=0.9$.  The difference between the upper and lower curves illustrates the frequency dependence of the relation between fitness and reproductive success.  Note that there is little curvature when an allele is rare, which leads to an advantage for rare types.  Redrawn from \textcite{frank90evolution}.}
\label{fig:fitness}
\end{figure}

\Fig{fitness} illustrates the two key properties of fitness.  First, fitness increases at a diminishing rate with a rise in reproductive success \autocite{gillespie77natural,frank90evolution}.  Put another way, the fact that fitness is a relative measure means that linear changes in reproductive success translate into nonlinear changes in fitness. 

Second, the curvature of the relation between reproductive success and fitness is frequency dependent \autocite{frank90evolution}.  A rare type has a nearly linear relation between reproduction and fitness.  A common type has a very strongly diminishing relation between reproduction and fitness.  This means that rare and common types are influenced differently by the consequences of variability, because more strongly diminishing returns cause variability in reproductive success to impose a greater penalty on fitness.  

\begin{figure}[t]
\includegraphics[width=0.85\hsize]{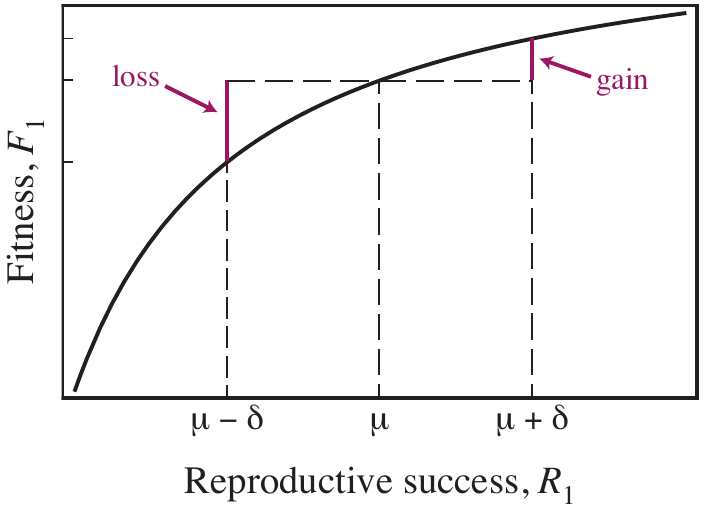}
\caption{Increasing variation in reproductive success reduces fitness.  Expected reproductive success is $\Gm$. Fluctuations of $\pm\Gd$ occur.  Positive fluctuations return a smaller gain in fitness than the loss suffered from a negative fluctuation.  Thus, equally frequent positive and negative fluctuations return a net loss. Redrawn from \textcite{frank90evolution}.}
\label{fig:diminish}
\end{figure}

\Fig{diminish} shows that diminishing returns cause a loss of fitness.  In the figure, expected reproductive success, $R$, is $\Gm$.  Deviations of $\pm\Gd$ occur, with increases and decreases at equal frequencies. The gain in relative fitness, $F$, for an increase of $\Gd$ units of reproductive success is less than the corresponding loss in fitness when reproductive success is reduced by $\Gd$. Expected fitness therefore declines as the frequency and magnitude of deviations increase. Note that the discount to fitness rises as the curvature between fitness and reproductive success increases.

\section{Reproduction multiplies and variation reduces success}

Suppose an individual has two offspring. Each of those offspring has two offspring.  The original individual has four grandchildren.  Compare that output to a second individual that has three offspring, and each of those offspring has one offspring. The second individual has three grandchildren.  In each case, the average reproduction per generation is two offspring.  However, the individual with less variable reproduction has greater success than the individual with more variable reproduction.

The difference occurs because reproduction multiplies over time.  The value $2\times2=4$ is greater than $3\times1=3$, even though the arithmetic averages are the same in each case.  In multiplicative series, variation reduces the multiplicative product.  Rather than measuring success by the arithmetic average, such as $(3+1)/2=2$, the proper average in a multiplicative series is the number that, when used to multiply in each generation, gives the total output.  In the reproductive series over two generations of $3\times1=3$, we need a number that when multiplied by itself gives three.  This works out as $\sqrt{3}\times\sqrt{3}=3$, so the multiplicative mean is $\sqrt{3}\approx 1.73$.  The multiplicative mean is usually called the \textit{geometric mean.}

\subsection{Approximation for the geometric mean}

A simple approximation of the geometric mean is often useful.  Suppose the arithmetic mean of a series is $\Gm$, and the variance of the series is $\Gs^2$.  Then the geometric mean is approximately $\Gm-\Gs^2/2\Gm$.  This approximation shows clearly how the variance reduces the geometric mean.

For example, in the series $3 \times 1 = 3$, the mean is $2$, the variance is $1$, the true geometric mean is $1.73$, and the approximation gives the geometric mean as $1.75$.  The smaller the variance is in relation to the mean, the better the approximation.  In evolutionary models, one often makes the assumption that average reproductive success is close to one, $\Gm\approx1$, so that the approximation for the geometric mean often appears as $\Gm-\Gs^2/2$.  

\subsection{The geometric mean principle}

Reproduction multiplies.  The greatest multiplicative series has the highest geometric mean reproduction.  Thus, the type with greatest long term fitness would appear to be the type with the highest geometric mean reproductive success.  That conclusion is often called the geometric mean principle. 

The highest geometric mean is sometimes associated with greatest evolutionary success. However, geometric mean reproductive success by itself often misleads, because it hides more than it helps.  The problem of evolutionary success and fitness turns out to be more subtle and more interesting.  The following sections explain why.

\section{Population regulation and relative success}

The total resources available to the population limit reproductive success.  That density dependent competition causes the reproductive success of each type to be influenced by the reproduction of other types.  For that reason, one cannot simply multiply the reproductive successes of each type independently and then compare the long-term geometric means.  Instead, each bout of density dependent competition causes interactions between the competing types.  Those interactions depend on frequency.  Reproduction of a rare type has little competitive effect on a common type.  Reproduction of a common type has a strong competitive effect on a rare type (\Fig{fitness}).

The fitness measure of relative success in \Eq{fitness} accounts for density dependent interactions.  The particular way in which density dependent competition arises has important consequences.  

\section{Expected change in frequency}

We can circumscribe the main conceptual issues by focusing on the expected (average) change in allele frequency.  Each process can be studied with respect to whether it tends to increase or decrease the expected frequency change.  

\Eq{freq} gives the definition for the allele frequency in the following generation, $q_1'=q_1F_1$.  The change in allele frequency is $\GD q_1=q_1'-q_1= q_1(F_1-1)$.  Using the definition for fitness in \Eq{fitness} allows us to write the change in frequency as
\begin{equation}\label{eq:freqChange}
  \GD q_1= q_1q_2\left(\frac{R_1-R_2}{\Rbar}\right).
\end{equation}
The reproductive successes fluctuate randomly.  Because those random fluctuations occur in both the numerator and the denominator, there is no simple way to express the exact change in frequency.  If we assume that the fluctuations in success are small relative to the average success, and we normalize all of the successes so that they are close to one, then we can write the approximate expected change in frequency as
\begin{equation}\label{eq:expChange}
  \E(\GD q_1) \approx q_1q_2\left\{(\Gm_1-\Gm_2)+\left[\cov(R_2,\Rbar)
  -\cov(R_1,\Rbar)\right]\right\},
\end{equation}
where $\Gm_1$ and $\Gm_2$ are the expected reproductive successes for types 1 and 2, respectively.  This equation, from \textcite{frank90evolution}, is equivalent to an approximation given by \textcite{gillespie77natural}.

The expected change is the average tendency.  Because of the inherent fluctuations in success, the actual change in frequency in any generation may be in the opposite direction of the expected change.  Over long time periods, three different patterns of evolutionary dynamics occur.

First, if the random fluctuations in the average reproductive successes of the types are large relative to the directional bias in \Eq{expChange}, then randomness dominates. The type frequencies will bounce up and down in a nearly neutral way.  Eventually, one type will become fixed and the other will disappear from the population.  If we start at frequency $q_1$, then the probability that type $A_1$ fixes is $q_1$, and the probability that type $A_2$ fixes is $q_2=1-q_1$. True fixation only occurs in finite populations.  In infinite populations, the related notion of quasi-fixation arises as the frequency of a type becomes very close to one.  To keep things simple, I ignore important technical distinctions between finite and infinite populations \autocite{gillespie94the-causes,ewens10mathematical}.

Second, if the directional bias is much larger than the fluctuations in the average reproductive successes of the types, then the type frequencies change in an almost deterministic way.  If the direction of change remains the same across the range of type frequencies, then the type with the greater expected success will usually become fixed in a relatively short period of time.  If frequency dependence provides a sufficient advantage to the rare type, then the direction of change may shift with type frequency in a way that tends to maintain both types in the population.

Third, if random fluctuations are of roughly the same magnitude as the directional bias, then frequency changes combine both directional and random aspects.  In some cases, frequencies will fluctuate, and both types will remain in the population for a very long time.  In other cases, frequencies may fluctuate for a significant period of time, but eventually one type or the other will become fixed.  Fixation will be biased toward the type favored by the directional tendency set by \Eq{expChange}.  However, the other type may occasionally fix because of the random fluctuations.  

I emphasize the major processes that influence the directional tendency in \Eq{expChange}.  In particular, the hierarchical structure of variability sets the directional tendency, which in turn shapes the qualitative patterns of dynamics.  I make only brief comments on long-term evolutionary dynamics, which require technical analysis of mathematical models to evaluate fully \autocite{gillespie94the-causes,ewens10mathematical}. As noted in the previous paragraphs, long-term dynamics include issues such as the probability that a particular genotype becomes fixed and the maintenance of polymorphism.  

\section{Hierarchical structure of variability}

Populations have a naturally nested hierarchical structure when considering genetics. Populations are composed of genotypes, and genotypes are composed of individuals.  In a haploid model with two alternative alleles, there are two genotypes in the population, and numerous copies (individuals) of each allelic type.  The hierarchical structure of variability makes explicit the variances and the correlations at different levels in the hierarchy.

For example, the reproductive success $R_1$ is the average success taken over all copies of the allele $A_1$. Similarly, $R_2$ is the average over all copies of $A_2$.  To analyze the variability in the aggregate success of allele $A_1$, one must consider the variability in the success of each copy of $A_1$ and the correlations in success between different copies.  The analysis for $A_2$ must consider the variability in the success of each allelic copy and the correlations in success between the alleles. 

We can relate the variances of individual reproductive success to the variances and covariances for the different allelic types by writing
\begin{subequations}\label{eq:corr}
	\begin{align}
	  \var(R_1)&=\Gr_1\Gs^2_1 \label{eq:a:corr}\\
	  \var(R_2)&=\Gr_2\Gs^2_2 \label{eq:b:corr}\\
	  \cov(R_1,R_2)&=\Gr_{12}\Gs_1\Gs_2\label{eq:c:corr}
	\end{align}
\end{subequations}
where $\Gr_1$, $\Gr_2$, and $\Gr_{12}$ are the correlations in reproductive success between randomly chosen pairs of $A_1$, $A_2$, or $A_1$ and $A_2$ individuals, respectively.   

\textcite{frank90evolution} introduced this explicit partitioning of variances and covariances for types into their individual components.  Any realistic analysis of variability must make explicit the individual level fluctuations and the associations between individuals.  Although this explicit treatment of variability is fundamental, the partitioning of variances and covariances in \Eq{corr} has often been regarded as some sort of highly technical or specialized analysis.  This mistake has limited progress in understanding fitness with respect to spatial and temporal fluctuations in success.

If we combine \Eq{expChange} and \Eq{corr}, and keep things simple by assuming that the correlation between types is zero, $\Gr_{12}=0$, we obtain
\begin{equation}\label{eq:expChange2}
  \E(\GD q_1) \approx q_1q_2\left\{(\Gm_1-q_1\Gr_1\Gs_1^2) - 
  	(\Gm_2-q_2\Gr_2\Gs_2^2)\right\},
\end{equation}
which means that, on average, type $A_1$ increases in frequency when
\begin{equation}\label{eq:expChange3}
  \Gm_1-q_1\Gr_1\Gs_1^2 > \Gm_2-q_2\Gr_2\Gs_2^2.
\end{equation}

The following sections show that different kinds of variability can be understood by the hierarchical partitioning of associations between traits within an individual and associations between different individuals. Variability interacts with the processes of density dependent population regulation.

\section{Temporal variability}

\textcite{dempster1955maintenance} introduced a model of temporal variation in which all alleles of the same type have the identical reproductive success within a generation, $\Gr_1=\Gr_2=1$, and there is no correlation between types, $\Gr_{12}=0$. In this haploid model, each individual has one allele, either $A_1$ or $A_2$.  The condition for the expected increase of type 1 from \Eq{expChange3} is
\begin{equation}\label{eq:temporal}
  \Gm_1-q_1\Gs_1^2 > \Gm_2-q_2\Gs_2^2.
\end{equation}
This equation illustrates the rare-type advantage induced by density dependent population regulation.  When the frequencies are equal, $q_1=q_2=1/2$, then the condition favors the type with the higher geometric mean fitness, $\Gm_1-\Gs_1^2/2$ versus $\Gm_2-\Gs_2^2/2$.  

As the frequencies approach one boundary, $q_1\rightarrow0$ and $q_2\rightarrow1$, the condition to favor $A_1$ becomes $\Gm_1 > \Gm_2-\Gs_2^2$.  At the other boundary, the condition favoring $A_1$ becomes $\Gm_1-\Gs_1^2 > \Gm_2$.  Thus, the directional tendency often shifts with frequency.  

In spite of the inherent rare type advantage, polymorphism is not maintained in this haploid model \autocite{gillespie73polymorphism,hartl73balanced,karlin74random}. The high variance in fluctuations eventually causes one of the types to fix (or to become nearly fixed in an infinite population).  The type with the higher geometric mean success has the advantage at the frequency midpoint.  That type fixes with higher probability.  If the geometric means for the two types are close to each other, then frequencies may fluctuate for a long time, and the bias toward fixing the favored type is relatively weak.  If the geometric means for the two types are significantly different, then fixation happens sooner and with a stronger bias toward the favored type.

\section{Correlations and genotypic homeostasis}

In the previous section, I assumed that all individuals carrying the same allele have the same reproductive success in each generation.  In that case, all of the variation arises from the response of an allele to environmental fluctuations, with no variation between individuals of the same genotype.  No variation means that individuals of the same type are perfectly correlated, $\Gr_1=\Gr_2=1$.  

Alternatively, different individuals of the same type may respond differently to environmental fluctuations.  There are many ways to express individual variation.  For example, individual responses may fluctuate about the long term arithmetic mean, $\Gm$, and the pairwise correlation between individuals in each generation may be $\Gr$ \autocite{frank90evolution}.  In that case, the variance in the average reproductive success of $A_1$ is $\Gr_1\Gs^2$, with a similar expression for $A_2$.  

Reduced correlation between individuals lowers the variation in the average success of a type.  That relation arises from the fact that the variance of an average is reduced by the number of uncorrelated observations in the sample.  We can express the effective sample size of uncorrelated observations as $n^*=1/\Gr$, so that the variance of the mean, $\Gs^2/n^*$, is $\Gr\Gs^2$. 

One can think of the pairwise correlations between individuals of the same genotype as the \textit{genotypic homeostasis}.  If all individuals of a genotype respond in exactly the same way to each environmental state, then the correlation between pairs of individuals is perfect, $\Gr=1$.  That perfect correlation increases the variance in the average reproductive success of the genotype.  One can think of such strong correlation as strong homeostasis or canalization of development for the genotype.  By contrast, weak correlation between individuals of the same genotype, with low values of $\Gr$, corresponds to greater developmental fluctuations and relatively weaker genotypic homeostasis.  

Given the variances in the average reproductive successes of the types as $\Gr\Gs^2$, the condition for $A_1$ to be favored is given in \Eq{expChange3}.  In a haploid model, the long-term bias in fixation depends on the relative geometric means derived when frequencies are equal, $q_1=q_2=1/2$, yielding the condition for $A_1$ to be favored as
\begin{equation}\label{eq:homeostasis}
  \Gm_1-\Gr_1\Gs_1^2/2 > \Gm_2-\Gr_2\Gs_2^2/2.
\end{equation}
This expression shows that temporal fluctuations favor reduced genotypic homeostasis, with low values of $\Gr$.  A type with low average reproductive success, $\Gm$, can be favored if it also has low genotypic homeostasis, $\Gr$, reducing its variance in average reproductive success sufficiently to give it a higher geometric mean fitness than its competitor.  Reduced genotypic homeostasis is a general expression of the widely discussed problem of bet hedging (see \Boxx{optimal}).

\section{Developmental variability}

\textcite{gillespie74natural} introduced a model in which the reproductive success of each of the $N$ haploid individuals in the population depends on its interactions with the environment during development. The reproductive successes of different individuals are independent because, by Gillespie's assumptions, different individuals experience different conditions and develop in an uncorrelated way. Nevertheless, the finite population size ensures that an individual's reproductive success correlates with the average reproductive success of its genotype. 

The correlation of two randomly chosen $A_1$ alleles is $\Gr_1 = 1/(Nq_1)$, because there are $Nq_1$ individuals of type $A_1$, and hence a chance $1/(Nq_1)$ of choosing the same individual twice. By the same reasoning, $\Gr_2= 1/(Nq_2)$. The correlation between types is zero, because different individuals experience different conditions. Substituting these values into \Eq{temporal}, we find that $A_1$ increases for any allele frequency when
\begin{equation}
  \Gm_1-\Gs^2_1/N > \Gm_2-\Gs^2_2/N.
\end{equation}
Since this condition no longer depends on allele frequencies, it is sufficient to describe long-term evolutionary

\begin{figure}[H]
\begin{minipage}{\hsize}
\parindent=15pt
\noterule
{\bf \noindent\BoxLabel. Optimal phenotypes in response to environmental variability}
\noterule
Individuals may be able to match their phenotype to particular environments.  Phenotypic plasticity occurs when an organism can sense the particular environmental state and adjust its traits accordingly.  If organisms do not adjust their phenotypes in response to the particular environmental state, then they may produce a stochastic response tuned to the pattern of fluctuation.  A stochastic response is sometimes called \textit{bet hedging.}

Bet hedging can increase the aggregate success of a genotype or strategy.  Suppose, for example, that the environment is equally likely to be in one of two states.  For each state, there is a different optimal phenotypic response.  However, the organism cannot adjust its phenotype in response to the particular state.  If each individual of the genotype has a random component to its phenotypic expression, then in each generation some individuals will match the environment with the best phenotype and some will not.  The mixture of phenotypic expressions reduces the variance of the aggregate success of the genotype by reducing the correlation between individuals of that genotype.  

The concept of reduced correlation between individuals of a genotype highlights an essential aspect of the bet hedging problem.  \textcite{frank90evolution} emphasized the general point about correlations between members of a genotype, as shown in \Eq{corr}.  \textcite{mcnamara95implicit} independently described a similar interpretation, but referred to the process as kin selection, a label that I would avoid in this case.  Correlations between relatives do matter, but not in the way that one usually associates with the costs and benefits of social behavior in kin selection.

Bet hedging can also arise within an individual. For example, the individual's alternative traits may be expressed stochastically in separate bouts of resource acquisition, in a way that reduces the overall variance in success.  Reduced variance in resource acquisition typically provides increased expected reproductive success, because success rises in a diminishing way with resources.

To get started on the literature, here are a few recent overviews for phenotypic plasticity \autocite{stearns89the-evolutionary,houston92phenotypic,moran92the-evolutionary,scheiner93genetics,via95adaptive,pigliucci01phenotypic,west-eberhard03developmental,dewitt04phenotypic} and bet hedging \autocite{cooper82adaptive,seger87what,sasaki95the-evolutionarily,grafen99formal,wilbur06life-history,king07the-evolution,donaldson-matasci08phenotypic}.  An interesting information theory approach may provide a useful connection between these subjects \autocite{kussell05phenotypic}.

All of these cases analyze phenotypic adjustment or phenotypic stochasticity. In these cases, one must also account for the diminishing relation between reproductive success and fitness (\Fig{fitness}) and the ways in which the correlations between individuals determine the mean and variance of aggregate success for each type (Eq.~\ref{eq:corr}).
\noterule
\end{minipage}
\end{figure}
\boxlabel{optimal}

\noindent advantage without the need to consider frequency dependence. An allele with a long-term advantage is more likely to become fixed than a neutral allele with the same initial gene frequency.

\textcite{gillespie74natural} presented this model of individual developmental variation as a separate problem from the general analysis of fluctuations in reproductive success.  The analysis here, from \textcite{frank90evolution}, shows that individual variation is just a special case of the general model of temporal variation.  One obtains the case of individual variation by properly calculating the correlations in reproductive success between individuals.

\section{Spatial variability and local population regulation}

The classical \textcite{dempster1955maintenance} model of temporal variation assumes that density dependent regulation occurs in one large population.  In that model, density regulation induces frequency dependence that favors the rare genotype.  I mentioned earlier that, in spite of the rare type advantage, one of the types eventually becomes fixed, because the random fluctuations in frequency are too strong relative to the directional tendency of evolutionary change. Fixation is biased toward the type with the highest geometric mean.

In a different model, \textcite{levene53genetic} showed that spatial variation does maintain genetic polymorphism.  In the Levene model, there are many independent spatial locations.  Each location has its own independent density dependent competition for resources.  

\textcite{gillespie74polymorphism,gillespie78a-general} showed that one can think of the Levene model of spatial variation as the sum of $K$ independent models of temporal variation.  If there is only one patch, $K=1$, then reproductive successes fluctuate over time in that patch, and all competition occurs in that single patch.  This model is identical to the classical Dempster model for temporal variation.

As $K$ increases, each independent patch fluctuates with the same rare type advantage of the classical Dempster model.  The total fluctuation in each generation is the average of the fluctuations over all patches.  Because the patches fluctuate independently, the variance of the average fluctuation over the entire population is reduced by $1/K$.  This reduction arises because the variance of the mean for $K$ independent observations is the variance of each observation divided by $K$.  

\textcite{gillespie74polymorphism,gillespie78a-general} provided the full analysis for this averaging over $K$ patches. However, he did not provide a simple interpretation or a simple expression for how fluctuations lead to a particular level of polymorphism when the number of patches is large.

\textcite{frank90evolution} noted that, as $K$ becomes large, the population-wide fluctuations in each generation become small because of the averaging effect over the many patches.  Thus, we can treat \Eq{expChange2} as an essentially deterministic process.  The rare-type frequency dependence now dominates.  The equilibrium frequency of types can be obtained by solving $\E(\GD q_1)=0$, which yields
\begin{equation}
  \frac{q_1}{q_2} = \frac{\Gm_1-\Gm_2+\Gr_2\Gs^2_2}{\Gm_2-\Gm_1+\Gr_1\Gs^2_1}
\end{equation}
as given by \textcite{frank90evolution}.  Here, each $\Gr$ is the correlation between copies of an allelic type measured within each patch. This result shows that naive comparison of geometric mean success is not sufficient to understand evolutionary outcome.

\section{Trait variability within individuals}

The theory above takes an individual's average and variance in reproductive success as given parameters.  However, the actual processes that lead to individual means and variances arise from the way that individuals acquire resources and produce offspring, including acquisition of food, protection from predators, and so on.  To analyze the full hierarchy of causes for variability, we should begin with the question:  How does the allocation of an individual's resources among alternative traits influence that individual's mean and variance in reproductive success?  

I use the example of traits for resource acquisition.  The same analysis applies to any trait that influences reproductive success, such as defense against parasites.

\subsection{One trait}

Let us start with a single trait.  The return of resources on investment has a random component, $\Gd$. The random component of resource acquisition affects reproductive success by an amount $f(\Gd)$.  Then a simple way to write the reproductive success is
\begin{equation}
  R=1+f(\Gd).
\end{equation}
If we assume that the random fluctuations, $\Gd$, have a mean of zero and a variance of $V_x$, and that the fluctuations are relatively small, then the average reproductive success is approximately
\begin{equation}
  \Gm \approx 1+f''V_x/2,
\end{equation}
where $f''$ is the second derivative of $f$ evaluated at zero \autocite{real80fitness,stephens86foraging}.  Typically, one assumes that fluctuations in traits for resource acquisition have a diminishing return shape as in \Fig{diminish}, in which case $f''<0$. Thus, greater fluctuations, $V_x$, reduce expected reproductive success.  All else equal, resource acquisition strategies with less variability yield higher average reproductive success than those strategies with more variability.  

The variance in an individual's reproductive success is approximately
\begin{equation}\label{eq:oneTraitVar}
  \Gs^2 \approx \var(f'\Gd) = f'^2 V_x,
\end{equation}
where $f'$ is the first derivative of $f$ evaluated at zero.

A full evaluation of fitness requires specifying the means, variances, and correlations between all individuals in the population.  The correlations must be evaluated in relation to the heritable types we are following over time.  The earlier sections provided the methods for studying evolutionary dynamics in relation to fitness.  

Here, to keep the focus on trait variability within individuals, I give only the geometric mean reproductive success for an individual. The geometric mean for each individual accounts for the average and variance in individual reproductive success but neither the correlations between types nor the role of density dependence in fitness. Assuming that the fluctuations in returns, $V_x$, are relatively small, the geometric mean reproductive success is approximately
\begin{equation}\label{eq:oneTrait}
  G=\Gm-\Gs^2/2\Gm \approx 1+\left(f''-f'^2\right)V_x/2.
\end{equation}

\subsection{Two traits}

How should an individual invest in two different traits that provide additive returns?  Let reproductive success be
\begin{equation}\label{eq:twoTraitsR}
  R=x\left[1+f(\Gd)\right] + y\left[(1-\Gg)+g(\Ge)\right]
\end{equation}
for investment amounts $x+y=1$, with $x$ and $y$ the fractions of total resources invested in each trait.  Here, $\Gg$ is the discount in expected return for the second trait, and $\Ge$ is the random fluctuation associated with the second trait.  The discount, $\Gg$, and the fluctuation, $\Ge$, are small relative to the baseline return of one. The mean of $\Ge$ is zero, and the variance is $V_y$.  I assume that $\Gd$ and $\Ge$ are uncorrelated.  \Boxx{twotraits} lists some of the intermediate steps.  Here, I focus on the key result.

If we assume that the random component of each trait is the same, $V_x=V_y$, and $f\equiv g$, then the geometric mean is
\begin{equation}\label{eq:twoTraitsGM}
  \Gm-\Gs^2/2\Gm \approx G + B(x,y),
\end{equation}
where $G$ is the geometric mean given in \Eq{oneTrait} for allocating all resources to the first trait, $x=1$, and $B(x,y)$ is the benefit obtained when mixing allocation of resources between the two traits such that $x+y=1$, with 
\begin{equation}
  B(x,y) = f'^2\left[1-(x^2+y^2)\right]V_x/2 -y\Gg.
\end{equation}
If we optimize $B$ to obtain the best mixture of allocations between the two traits, we obtain
\begin{subequations}\label{eq:twoTraitsOpt}
	\begin{align}
	  x^* &= \frac{1}{2}\left(1 + \frac{\Gg}{\Gs^2}\right) \label{eq:a:twoTraitsOpt}\\[5pt]
	  y^* &= \frac{1}{2}\left(1 - \frac{\Gg}{\Gs^2}\right), \label{eq:b:twoTraitsOpt}
	\end{align}
\end{subequations}
where $\Gg$ is the discount in expected return for the second trait given in \Eq{twoTraitsR}, and $\Gs^2$ is the variance in individual reproductive success per trait given in \Eq{oneTraitVar}.  

\begin{figure}[H]
\begin{minipage}{\hsize}
\parindent=15pt
\noterule
{\bf \noindent\BoxLabel. Trait variability with two traits}
\noterule
This box shows the details that lead from \Eq{twoTraitsR} to \Eq{twoTraitsGM}.  Starting with \Eq{twoTraitsR} for $R$, the average reproductive success is approximately
\begin{equation}
  \Gm \approx 1-y\Gg +xf''V_x/2+ yg''V_y/2.
\end{equation}
The variance in success is approximately
\begin{align*}
  \Gs^2 &\approx \var(xf'\Gd+yg'\Ge)\\
        &=x^2 f'^2 V_x + y^2 g'^2 V_y.
\end{align*}
The geometric mean is approximately
\begin{equation*}
  \Gm-\Gs^2/2\Gm \approx 1-y\Gg + x\left(f''-xf'^2\right)V_x/2 + y\left(g''-yg'^2\right)V_y/2.
\end{equation*}
If we assume that the random component of the two traits is the same, $V_x=V_y$, and $f\equiv g$, then the geometric mean is
\begin{align*}
  \Gm-\Gs^2/2\Gm &\approx 1-y\Gg + f''V_x/2-(x^2+y^2)f'^2V_x/2\\
   &= 1 + \left(f''-f'^2\right)V_x/2 \\
        &\mskip100mu + f'^2\left[1-(x^2+y^2)\right]V_x/2 -y\Gg.
\end{align*}
The substitutions given in the text lead directly to \Eq{twoTraitsGM}.
\noterule
\end{minipage}
\end{figure}
\boxlabel{twotraits}

It pays to invest some resources in the trait with lower expected return as long as $\Gg/\Gs^2 < 1$.  The lower expected return is offset by the reduced variance obtained from averaging the returns over two uncorrelated traits.  In both biology and financial investing, returns tend to multiply over time.  Thus, reduced fluctuations enhance the multiplicative (geometric) average return.  In financial investing, the central role of the geometric mean is well known in theory \autocite{bernstein97diversification}, but often ignored in practice \autocite{macbeth95whats}.

\subsection{An example}

The concepts in the previous section are simple.  The variance of an average declines with additional uncorrelated components.  Reduced variance provides a benefit when success multiplies over time.  The technical expressions of those results may obscure the simplicity of the concepts.  This section provides a numerical example.

Suppose an organism has two different behaviors by which it can obtain calories.  To keep the problem simple, assume that there is a linear relation between calories and reproduction, $f'=1$.  For the first behavior, the return is on average $\Gm_1=1.0$ calories, with a variance in return of $\Gs^2=0.1$.  The second behavior has a lower average return of $\Gm_2=1.0-0.02=0.98$ calories, with the same variance of $\Gs^2=0.1$.  

If all investment is devoted to the first behavior, then the geometric mean success is $\Gm_1-\Gs^2/2=0.95$.  If all investment is devoted to the second behavior, then the geometric mean success is $\Gm_2-\Gs^2/2=0.93$.  

In this case, I have assumed $\Gg=0.02$ and $\Gs^2=0.1$. From \Eq{twoTraitsOpt}, the optimal allocation to the two traits is $x^*=0.6$ and $y^*=0.4$.  If the individual devotes a fraction $0.6$ of its investment to the first behavior and a fraction $0.4$ of its investment to the second behavior, then it obtains an average return of 
\begin{equation}
  a=0.6\Gm_1+0.4\Gm_2=0.992.
\end{equation}
The variance in return is obtained by noting that, when one splits allocation between two uncorrelated returns, $R_1$ and $R_2$, each with variance $\Gs^2$, the variance is 
\begin{equation}
  b=\var(xR_1+yR_2) = [x^2+y^2]\Gs^2.
\end{equation}
Using the optimal split $0.6$ versus $0.4$ for $x$ and $y$, and the value $\Gs^2=0.1$ above, the variance is $b=0.052$.  The geometric mean is now approximately
\begin{equation}
  a-b/2=0.992-0.052/2=0.966.
\end{equation}
This mixture of behaviors therefore returns a higher geometric mean of $0.966$ than when all investment is devoted to the higher yielding first behavior, which has a geometric mean of $0.95$, or when all investment is devoted to the lower yielding behavior, which has a geometric mean of $0.93$.  This example illustrates the benefit of diversification when success multiplies over time.  

Note that reproductive returns are linear in this example.  The entire benefit of diversification arises from the multiplicative nature of long-term success, which discounts variance.

\subsection{The limitation of using individual geometric mean success}

I used the individual's geometric mean success in this section.  That assumption is valid only when we are interested in an absolute measure of an individual's long-term success in the absence of competition and relative comparison with others.  However, it often does not make sense to measure success independently of others.  Evolutionary success depends on the relative contribution to the population by a heritable type.  The earlier sections of this paper showed several different measures of success that arise from the temporal and spatial structure of competition and from the correlations in success between different types of individuals.

Suppose, for example, that the correlation between individual copies of an allele is low, $\Gr \rightarrow 0$, as in the model of developmental variation in which $\Gr=1/N$ and population size, $N$, is large.  Then, from \Eq{homeostasis}, we see that natural selection favors the type with the highest arithmetic average, $\Gm$, independently of the individual variance in reproductive success, $\Gs^2$.  In that case, it does not make sense to analyze the geometric mean of individual reproductive success.  Instead, success depends almost entirely on the arithmetic mean return taken over the two traits. If returns per trait are linear, $f''=0$, then the arithmetic mean is
\begin{equation}
  x\Gm_1+y\Gm_2=x + y(1-\Gg).
\end{equation}
In this case, individuals will be favored to allocate all resources to the higher yielding trait, labeled as trait one in this example.

\section{Economic theories of variability and risk}

Economic theories of risk and uncertainty typically focus on the absolute success of individuals or single agents.  Relative success in economics concerns market share \autocite{frank90when}.  However, there seems to be little economic theory about risk and uncertainty in relation to market share.  Problems of market share lead to many issues discussed in this paper.  For example, relative success induces diminishing returns.  The temporal and spatial scale of competition determines the proper measure of success.  

One must also consider the proper unit of analysis to measure success and dynamics.  If one is interested in the absolute currency value accumulated by an individual investor over a long period of time, then the individual's geometric mean success in return over successive intervals is often a good measure.  If one is interested in an individual's purchasing power, then one must track the individual's currency valuation relative to the currency valuation among the population of individuals competing for the same goods.  

If the individual has only a small fraction of the total pool of goods, then the individual's geometric mean return provides a good measure of success.  However, if one is tracking a corporation or agent that controls a large fraction of the total resource pool, then the correlation between individual and total success may have a significant impact on outcome.

In some economic analyses, one is interested in behaviors or strategies.  For example, what is the relative success of those following a particular financial strategy in the investment markets?  The answer depends in part on whether all individuals following the same strategy have highly correlated returns or uncorrelated returns.  A high correlation in returns between individuals following a strategy increases the variance in the aggregate success of that strategy.  Higher variance usually leads to lower long-term success.  

The frequency of the competing strategies also matters.  Relatively rare strategies have a low correlation with the population average level of success, providing a rare-type advantage.  The consequences of the rare-type advantage depend on whether competition occurs globally or locally over a series of isolated markets. These problems of relative success have received little attention in the study of economic competition.

\begin{figure}[H]
\begin{minipage}{\hsize}
\parindent=15pt
\noterule
{\bf \noindent\BoxLabel. Reviews and technical issues}
\noterule
\textcite{gillespie94the-causes} and \textcite{ewens10mathematical} provide excellent technical overviews of genetic theory for variable environments.  Recent reviews \autocite{hedrick06genetic,proulx10the-standard} and new theory \autocite{taylor08environmental} continue to appear. 

\textcite{lande08adaptive} developed a comprehensive approach to the theory of fluctuating selection.  He emphasized an adaptive topography method arising from the key insight that expected fitness depends on average reproductive success minus the covariance of reproductive success with population mean success.  That expression for fitness is the same as developed in \Eq{expChange}, following from \textcite{gillespie77natural} and \textcite{frank90evolution}.  

\textcite{lande08adaptive} also summarized the complexities that arise in diploid genetic systems under fluctuating selection.  In general, diploid and multilocus models require one to pay attention to two issues \autocite{frank90evolution}.  First, how do the fluctuations contributed by different alleles combine within individuals to determine the average and variance in individual reproductive success?  Second, how do multiple alleles per individual induce correlations in reproductive success between copies of alleles in different individuals? The specific results that I gave in the text concern haploid models.  Diploid and multilocus models must account for these additional complexities.

\textcite{rice08a-stochastic} introduced an exact expression for evolutionary change in stochastic models by expanding the scope and interpretation of the Price Equation.  \textcite{tuljapurkar09from} review the theory of variable environments in relation to demography and life history evolution.  For economics theory, introductory microeconomics texts include overviews of the theory of risk and uncertainty.  For investments, see \textcite{markowitz91portfolio}.  \textcite{okasha11optimal} provides an entry to the philosophical literature on the theory of uncertainty in evolution and economics.
\noterule
\end{minipage}
\end{figure}
\boxlabel{reviews}

\section{Conclusions}

Nearly all aspects of success include a variable component.  To understand the consequences of that variability, one must study the hierarchical structure of traits within populations. Each individual has multiple traits.  Each genotype or strategy has multiple individuals.  Fluctuation in the success of particular traits has consequences that depend on the correlations between traits and the correlations between individuals.  The aggregate variability of competing types affects relative success in ways that depend on density dependent competition, which causes diminishing returns and induces an intrinsic frequency dependence that tends to favor rare types over common types.  

The extensive biological theory of variability has dealt with particular aspects of the overall problem.  But few analyses have set out the entire range of fluctuations, how those fluctuations are structured in populations, and the particular nature of competition that shapes the consequences of fluctuations.  By considering the structure of the entire problem, one obtains a richer understanding of biological fitness and its consequences for evolutionary dynamics.

Many of the biological problems of variability also arise in economics.  The theoretical literature in economics made the first analyses of success when there is a variable component of returns.  But the biological literature has advanced further in the analysis of variability, particularly with respect to the importance of relative success and the hierarchical structure of competing types or strategies in populations.  

\begin{acknowledgments}
I thank Sylvain Gandon for helpful comments. Parts of this article derive from \textcite{frank90evolution}. My research is supported by National Science Foundation grant EF-0822399, National Institute of General Medical Sciences MIDAS Program grant U01-GM-76499, and a grant from the James S.~McDonnell Foundation.  
\end{acknowledgments}

\bibliography{main}
\end{document}